\documentclass[12pt,twocolumn,preprint,a4]{iopart}

\usepackage{graphicx}
\usepackage{booktabs}
\usepackage{textcomp}

\begin{document}

\title[Scaling and laws of DC discharges as pointers for HiPIMS plasmas]
      {Scaling and laws of DC discharges as pointers for HiPIMS plasmas}

\author{C. Maszl$^{1}$\footnote{\ead{christian.maszl@rub.de}}, J. Laimer$^2$, A. von Keudell$^1$ and H. St\"ori$^2$ }

\address{$^1$Research Department Plasmas with Complex Interactions,
Ruhr-Universit\"at Bochum, Institute for Experimental Physics II,
D-44780 Bochum, Germany\\
$^2$Institute of Applied Physics, Vienna University of Technology, 1040 Vienna, Austria}

\date{\today}

\begin{abstract}
Scaling or smiliarity laws of plasmas are of interest if lab size plasma sources are to be scaled for industrial processes. In the case of direct current (DC) magnetron sputtering the scales can range from few centimeters in the lab to several meters if one takes the flat glass industry as one example. Ideally, the discharge parameters of the scaled plasmas are predictable and the fundamental physical processes are unaltered.  Naturally, there are limitations and ranges of validity. \\
Scaling laws for direct current glow discharges are well known. If a well diagnosed discharge is scaled, the field strength in the positive column, the gas amplification and the normal current density can easily be estimated. For non-stationary high power discharges like high power impulse magnetron sputtering (HiPIMS) plasmas, scaling is not as straight forward. Here, one deals with a non-stationary complex system with strong changes in plasma chemistry and symmetry breaks during the pulses. Because of the huge parameter space no good parameters are available to define these kind of discharges unambiguous at the moment.\\
In this contribution we will discuss the scaling laws for DC glow discharges briefly and present experimental results for a discharge with copper electrodes and helium as plasma forming gas. This discharge was operated in a pressure range from $200$ to $1600\;$hPa with three different electrode diameters ($D=2.1$, $3.0$ and $4.3\;$mm). Results from breakdown voltage measurements indicate that the pressure and electrode distance cannot be varied independently. Also limitations of the theoretical Paschen curves will be adressed.  Effects of the scaling on the reduced normal current density, the reduced normal electric field in the positive column and discharge temperature will be discussed and limitations highlighted.\\
Compared with these results the added complexity of HiPIMS plasmas will be described. Suggestions in the community how to obtain experimental finger prints of the plasmas will be reviewed and implications from the experiences with DC discharges on the development of similarity laws for HiPIMS plasmas will be discussed.
\end{abstract}

\maketitle

\section{Introduction}
In the past, technological plasma were developed without a sound physical understanding. During the last decades the situation improved a lot. Sophisticated diagnostics and simulation tools were developed to gain more insight into industrial processes and lead to a tremendous success of plasma technologies in the industry.\\
Still, a lot of developments in the thin film industry are achieved by an trial and error approach. This process is time consuming, expensive and it is often not clear whether the most efficient synthesis procedure is found or not. Knowledge based thin film design is therefore one of the key challenges in the future. Here one has to understand how certain thin film properties are connected to plasma parameters and how these are connected to the power supply of the setup. Furthermore, scaling of these systems is of importance. Labscale magnetrons have a typical size of a few cm where magnetrons in the flat glass industry can span several meters as one example. It is therefore important to find good reduced parameters which define a certain plasma unambiguously. These set of parameters are usually refered to as similarity or scaling laws.\\

For non-stationary high power plasmas these parameters are not available at the moment. High power impulse magnetron sputtering (HiPIMS) is an instructive example for these type of systems \cite{Kouznetsov1999,Sarakinos2010,Gudmundsson2012}.  HiPIMS plasmas are created during short high voltage pulses in the $\mu$s range on a magnetron. Peak power densities at the targets can reach several MW/m$^2$. These high powers cause high ionization degrees up to 90$\%$ of the sputtered species and huge thermal loads on the target. To avoid thermal damage the duty cycle is only a few $1\%$ to allow the target to cool down. The produced thin films have an improved adhesion, low surface roughness, are very dense and hard. One of the important features of HiPIMS is a strong change in plasma chemistry \cite{Hecimovic2012}. Typically, the discharge ignites in argon. Due to the strong sputterwind rarefaction sets in and the noble gas is replaced by sputtered species \cite{Hoffman1985,Rossnagel1998,Palmucci2013}. Depending on operation conditions the plasma can enter the runaway regime where the discharge current increases until the end of the pulse \cite{Anders2011}. Below the runaway regime the plasma current has a flat top after an initial current peak. During the pulses the plasma often shows self-organization and symmetry breaks. These are localized ionization zones and are known as spokes \cite{Ehiasarian2012,Winter2013,Brenning2013}. Their quasi-mode number is typically between one and four for a 2-inch magnetron. HiPIMS plasma are non-stationary complex systems and therefore hard to describe. Unambigous, reduced parameters are currently not available and are hard to deduce. A simple plasma where scaling laws are known are unmagnetized direct current (DC) glow discharges. In this contribution we will discuss these laws and speculate if they can act as a pathfinder for HiPIMS plasmas. 

\section{Scaling laws of direct current glow discharges}
\label{sec:scalinglaws}
Scaling laws predict the behaviour of a discharge when its parameters are changed and facilitate the discrimination between fundamental processes \cite{Engel1994}. For DC glow discharges usually a set of four equations \cite{Roth1995}  is used which are valid over a wide range of parameters \cite{Gambling1957}
\begin{equation}
pd=\textrm{const.} \qquad \frac{E}{p}=\textrm{const.} \qquad \frac{j}{p^2}=\textrm{const.} \qquad \frac{\alpha}{p}=\textrm{const.}
\label{equ:scaling_laws}
\end{equation}
These equations depend on the pressure $p$, the electrode distance $d$, the field strength in the positive column $E$, the normal current density $j$ and the gas amplification or first Townsend coefficient $\alpha$. This means as long as the product of pressure $p$ and the spatial coordinate $d$ is kept constant the scaled discharge will have the same discharge current and voltage drop for a certain working gas and electrode material. This is true as long as the temperature remains constant during scaling. Therefore the density $n$ would be a better parameter instead of the pressure. Since spatial resolved density or temperature values are hard to obtain for high pressure discharges, scalings with the pressure $p$ are used in this study. Additionally to the scaling laws, the cathode fall thickness $d_C$ and cathode fall voltage $V_C$ are often tabulated in textbooks like \cite{Raizer1991}.\\

\begin{figure}[ht]
	\centering
	\includegraphics[width=8.5cm]{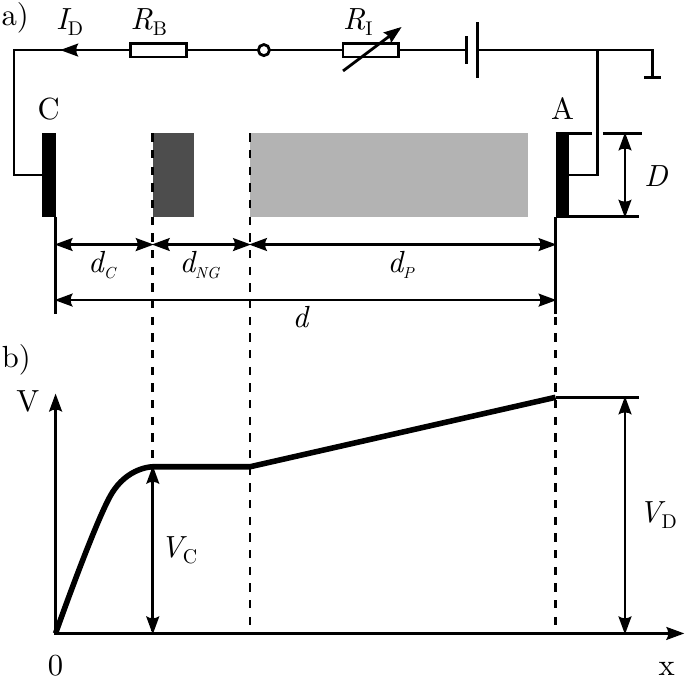}
	\caption{a) Experimental setup. $R_B$ ist the 3.3~k$\Omega$ ballast resistor. $R_I$ indicates the internal current limiter of the power supply. The negative glow with thickness $d_{NG}$ is indicated as dark gray and the positive column with a length $d_P$ as light gray area. All other patterns are omitted for simplicity b) Simplified potential profile of a DC glow discharge. The anode (A) fall is neglected in this representation. $V_C$ is the cathode (C) fall voltage with thickness $d_C$, $V_D$ the total discharge voltage and $I_D$ the current. All distances and potentials are not to scale. Partially reproduced from \cite{Piel2010}, p.~325}
	\label{fig:setup_profile}
\end{figure}

As a case study of these parameters, a DC glow discharge (Fig.~\ref{fig:setup_profile}a) will be designed in the following: Anode (A) is grounded. The cathode (C) is connected to a power supply with an internal current limiter $R_I$. Additionally there is a ballast resistor $R_B$ to mitigate oscillations of the internal controller. The negative glow is indicated by a dark shaded area, the positive column with a light gray area.

In order to choose the right power supply, the discharge voltage $V_D$, the current $I_D$, the total power consumption and the necessary ignition voltage have to be calculated. In this example design goal are Fe electrodes with a radius $R=5$~mm and an inter-electrode gap of $d=10$~mm at a helium pressure of $p=320$~hPa. The product of pressure and radius gives a $pR$ value of 800~hPa$\,$mm and a $pd$ value of 3200~hPa$\,$mm. For simplicity one assumes a simplified quadratic potential profile in the cathode fall, a constant potential in the negative glow region and a linear increasing potential in the positive column (Fig.~\ref{fig:setup_profile}b). The anode fall which neglected. Typically, it is to the order of the first ionization potential of the working gas (Fig.~\ref{fig:setup_profile}b). The breakdown voltage is of the order of $1000$~V according to Fig.~\ref{fig:paschen}. The required scaling parameters for the design are found in \cite{Raizer1991} und summarized in Tab.~\ref{tab:reduced_param}.
\begin{table}[ht]
	\centering
	\begin{tabular}{|c|c|c|c|}
		\hline
		$(j/p^2)_n$ 	  & $V_C$ & $pd_C$ 	      & $(E/p)_n$ \\
		mA/(hPa$^2$m$^2$) &  V    & $10^{-3}\,$hPa$\,$m & V/(hPa$\,$m) \\ 		
		\hline
		12.4		  & 150	  & 17.3	      & $80$\\
		\hline
	\end{tabular}
	\caption{Reduced current density (\cite{Raizer1991}, p.~183), normal cathode fall voltage, reduced cathode fall thickness (p.~182) and normal reduced electric field  (p.~196) in the positive column for Fe electrodes and helium at room temperature.}
	\label{tab:reduced_param}
\end{table}

The discharge current $I_D$ is calculated by multiplying the normal reduced current density from Tab.~\ref{tab:reduced_param} by the electrode surface $A=D^2\pi/4$ and the pressure $p$ which yields a discharge current of $I_D=25$~mA. The discharge voltage $V_D$ depends on the potential difference in the positive column and the cathode fall voltage $V_C$. Since the electric field in the positive column is given by the scaling law (Tab.~\ref{tab:reduced_param}) the only missing quantity is the length $d_P$ of the positive column (Fig.~\ref{fig:setup_profile}a). The thickness of the cathode fall is $d_C=pd_C\,p=54$~$\mu$m. Usually, the combined thickness of the negative glow and the Faraday dark space $d_{NG}$ is not known. For high pressure discharges this length is orders of magnitude longer than $d_C$ (\cite{Maszl2011b}, cf. Fig.~1b). We therefore set the length of the positive column to the inter-electrode gap distance $d_P=d=10$~mm which is justified for long discharges like in our case. Since the potential in the $d_{NG}$ region is approximately constant \cite{Piel2010}, the potential difference in the positive column is  overestimated.\\
According to \cite{Raizer1991}, p.~196 for $pR>105$~hPa$\,$mm the  normal field strength in the positive column is approximately $(E/p)_n=0.075$~V/(hPa$\,$mm). The potential difference in the positive column is then $V_P=(E/p)_n\,p\,d_P=240$~V. Together with the cathode fall voltage $V_C$ this yields a discharge voltage of $V_D=390$~V. The total power consumption of the discharge is $P=V_DI_D\approx10$~W. Almost $4$~W are only dissipated in the thin cathode fall region. Summarising, one needs a power supply with a maximum voltage of approximately 1000~V to allow ignition (Fig.~\ref{fig:paschen}) and which is capable of providing $10$~W continuous power. The additional power consumption of the ballast resistor $R_B$ has also to be taken into account.\\

But how likely are these results? It is instructive to compute the power densities for the cathode fall volume ($P_C/V_C$) and the positive column volume ($P_P/V_P$) which give 3.5~kW/cm$^3$ respectively 0.03~kW/cm$^3$. These high power densities cause heating of the electrodes, the neutral gas and therefore affects the resistivity of the discharge. A density correction for the scaling laws was already proposed in 1933 by \cite{Engel1933}. For the derivation of the average temperature in the cathode fall region $\overline{T}$, A. von Engel et al. assume a linear problem, a linear cathode fall and that heating of the gas neutrals is caused by friction with ions. Additionally, the total produced heat by this process is absorbed by a cooled electrode at $T_{E}$ which gives at the end for the pressure $p$ depending average temperature in the cathode fall region $\overline{T}$
\begin{equation}
\overline{T}-T_{E}=\frac{2}{3}\sqrt{\frac{V_Cj_nd_n}{3\alpha_G}}\sqrt{\frac{p\;T_0}{\overline{T}}}.
\label{equ:tbar}
\end{equation}
$\alpha_G$ is a gas specific constant and is proportional to the thermal conductivity $\lambda$ over $T$. $j_n=(j/p^2)_n$ is the reduced normal current density and $d_n=pd_C$ the reduced normal cathode fall thickness at normal conditions. $T_E$ is the temperature of the coolant and $T_0$ the temperature at normal conditions. Results and a comparison to measured temperatures will be given in Sec.~\ref{sec:jp} and Sec.~\ref{sec:Ep}.\\

Apart from thermal effects scaling laws are also influenced by microscopic processes. A. v. Engel \cite{Engel1994} identified allowed and forbidden phenomena. Without a claim to be complete allowed processes are ionization by single collisions, ambipolar diffusion, electron attachment and detachment, neutralization charge transfer and ion-ion recombination at high pressures. A violation of the similarity laws occur if stepwise ionization, photoelectric and field emission, ionization charge transfer and ion-ion recombination at low pressures takes place.

\section{Experimental results and discussion for DC glow discharges}
Following the same logic like as for the case study in Sec.~\ref{sec:scalinglaws} we will first describe the ignition behaviour of the discharge, followed by the reduced normal current density and the field strength in the positive column. All measurements are discussed right after the experimental results.\\

\begin{figure}[ht]
	\centering
	\includegraphics[scale=1.00]{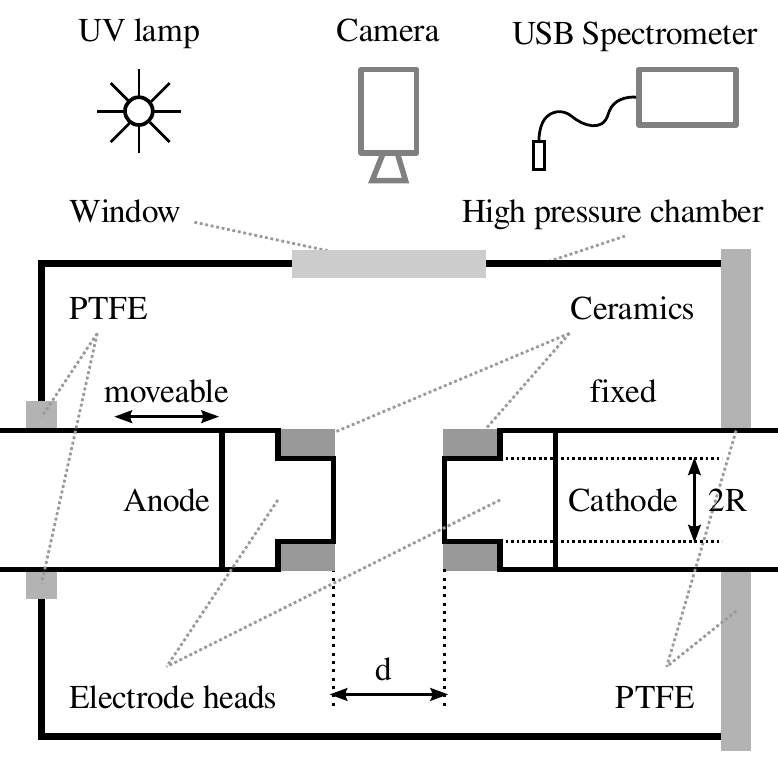}
	\caption{Schematic sideview of the high pressure chamber with a quartz glass window (not to scale). The anode is moveable. Ceramic tubes ensure planar electrodes. Both electrode heads can be changed to allow diameter variations. Standard diagnostics are a camera in combination with a microscope and a USB spectrometer. An UV lamp can be used to irradiate the electrodes to mitigate breakdown delay times.}
	\label{fig:DC_expsetup}
\end{figure}

The experiments were performed in a home-made high-pressure chamber with a pressure range from $0-2000$~hPa (Fig.~\ref{fig:DC_expsetup}). The pressure was adjusted with a valve in the gas feed and in the pump line and measured with a MKS Baratron and a WIKA EN837-1 bourdon tube for pressures above 1000~hPa. Helium 5.0 was the plasma forming gas. The setup allowed the variation of the interelectrode gap during operation between $0$ and $15$~mm with a resolution of $0.05$~mm. Furthermore, the water-cooled cylindrical copper electrodes could be replaced, thus allowing the variation of the electrode diameters between $2.1$ and $15$~mm. To allow measurements in the anormal glow regime the electrodes are shielded with ceramic tubes. In contrast to experiments in the low pressure regime the produced plasma is not bounded by the chamber wall but surrounded by neutral gas instead. The electrodes were actively cooled with a pipe in pipe cooling system to cope with an increase in the power density during downscaling.\\

A current-controlled high-voltage power supply unit (Fug MCN350-2000) and an additional $3.3$~k$\Omega$ ballast resistor was used to set the operating point of the discharge. Spectra were taken with an Ocean Optics optical spectrometer S4000 and a HR4000 ADC. During all measurements OH was present as impurity. A fit of the OH-band with Lifbase \cite{Luque1999} gave the rotational temperature as an upper boundary for the gas kinetic temperature. Conventional multimeters (Fluke 75 Serie II) were used for current and voltage measurements. An Olympus stereo microscope (SZX12) in combination with an Olympus 330e consumer camera was used to take pictures of the discharge. The obtained pictures were processed with ImageJ \cite{Schneider2012} to obtain gray value profiles and from the full width half maximum an estimate for the diameter of the positve column.
\subsection{Breakdown voltages}
Cylindrical, polished copper electrodes with a diameter of 15~mm were used to measure breakdown voltages. These electrodes had no ceramics shield and the same diameter as the probe head holder (Fig.~\ref{fig:DC_expsetup}). The electrodes were irradiated with UV radiation from a mercury-vapor lamp to create seed electrons and avoid breakdown delay times \cite{Kudrle1999}. Two different experiments were conducted. The first was performed at a fixed interelectrode gap and a pressure variation according to the parameters in Tab.~\ref{tab:param_paschen}, no.~I. 
\begin{table}[htdp]
\caption{Electrode diameters, distance and pressure ranges for breakdown voltage measurements.}
\begin{center}
\begin{tabular}{|l|l|l|l|}
\hline
No. & Diameter & Distance & Pressure  \\
    & mm       & mm       & hPa\\
\hline
I)  & $15$ & $8.80$ & $10\to 1600$ \\
II) & $15$ & $0.25\to 8.00$ & $50, 200, 800, 1600$ \\
\hline
\end{tabular}
\end{center}
\label{tab:param_paschen}
\end{table}

\begin{figure}[ht]
	\centering
	\includegraphics[scale=1]{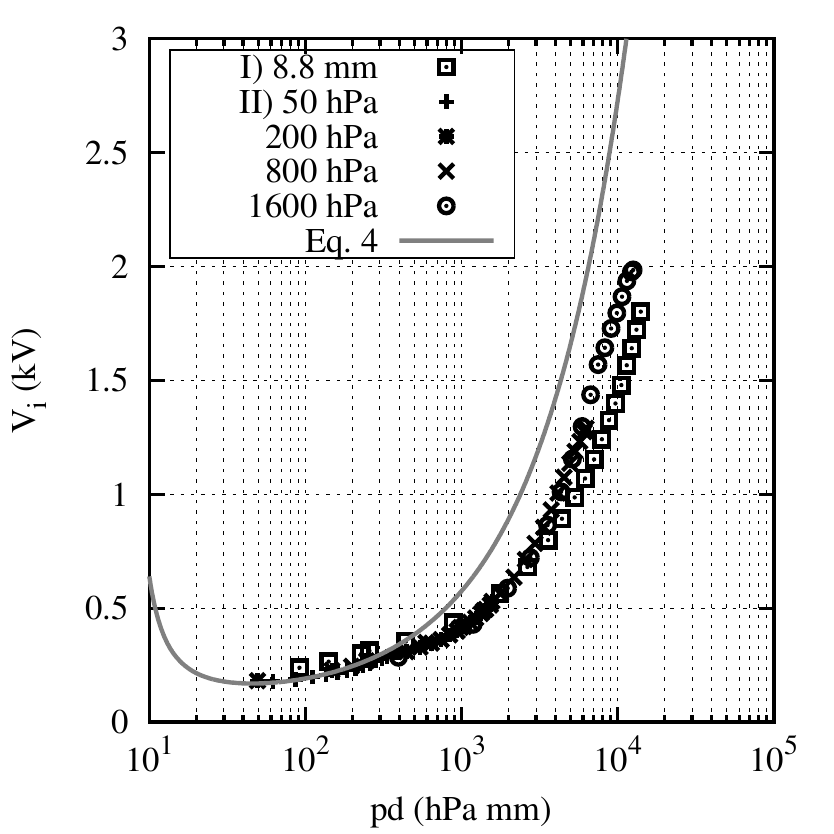}
	\caption{Theoretical Paschen curve (grey solid line) and curves for pressure ($p$) variation (white squares) and piecewise distance ($d$) variation.}
	\label{fig:paschen}
\end{figure}

In the second study II) the interelectrode gap was varied. This was necessary to cover the same $pd$ range as in experiment I. The variation was done therefore at four different pressures (Tab.~\ref{tab:param_paschen}, no.~II). The voltage was increased slowly ($\sim$V/s) and the breakdown voltage was directly read from the power supply and plotted in Fig.~\ref{fig:paschen}.

The measured values are compared to calculated breakdown voltages $V_i$ from the semi-empirical Paschen formula for inert gases (solid line)
\begin{equation}
V_i=\frac{B^2\;pd}{\ln^2(C)} \qquad C=\frac{\ln(1/\gamma+1)}{A\,pd}.
\label{equ:paschen}
\end{equation}
$A=0.33$~hPa$^{-1}$mm$^{-1}$ and $B=3.86$~V(hPa$\;$mm)$^{-1/2}$ are empirical values which are valid for a reduced field strength of $E/p=7.6$~V$\,$hPa$^{-1}$mm$^{-1}$. These two parameters are usually obtained by measuring the first Towsend coefficient $\alpha$ which is later used in the ignition condition to obtain the Paschen law. A total electron emission coefficient $\gamma$ of 0.15 was used to fit the experimental data for low $pd$ values. This is in a good agreement with other results from \cite{Raizer1991}, p.~134.
\subsubsection{Discussion}
Both measurements (Fig.~\ref{fig:paschen}) start to deviate from the theoretical curve around $pd\approx 500$~hPa$\;$mm. One has to keep in mind that the Paschen curve is a semi-empirical formula. The biggest uncertainty lies in the gas amplification coefficient $\alpha$. For low reduced field strengths this coefficient has large gradients and can yield strong variations depending on the quality of the fit or experimental results \cite{Maric2005,Kuschel2013}. The fit parameters $A$ and $B$ are here only valid for reduced field strengths around $E/p\approx 7.6$~V$\,$hPa$^{-1}$mm$^{-1}$. For high pressure discharges $E/p$ is typically rather low. In this study it ranges from 0.13 to 2.7~V$\,$hPa$^{-1}$mm$^{-1}$ which is clearly below the reduced field strength given above. Another source for lower breakdown voltages is field amplification at the edges of the cylindrical electrodes. This effect  was not compensated for the used electrodes. In order to achieve a reduction of the field near the edges Rogowski profiles \cite{Rogowski1923,Rogowski1926} can be machined to the electrodes.\\  

The Paschen law implies (Eq.~\ref{equ:paschen}) that the pressure $p$ and the distance $d$ can be varied independently. For values higher of $pd\approx 3000$~hPa$\;$mm the measured curves start to deviate from each other. Apparently, the aspect ratio $d/D$ of the discharge has some effect. Lisovskiy et al. \cite{Lisovskiy2000} proposed a modified fit formula for cylindrical electrodes which takes the aspect ratio into account. According to their result they claim that the original Paschen law is only valid for low aspect ratio discharges with $d/D\to0$. Furthermore, in their opinion each point measured by method Tab.~\ref{tab:param_paschen}, no.~I) represents a point on a different genuine Paschen curve.
\subsection{Reduced normal current density}
\label{sec:jp}
The reduced normal current density $j_n=(j/p^2)_n$ was determined during IV-characterisation of similar discharges. Tab.~\ref{tab:parameters_scaling} gives the parameters of the three measurement series. All copper electrodes were machined to fit in three different Al$_2$O$_3$ ceramic tubes to allow measurements in the anormal glow region. Otherwise the negative glow would start to cover the cylinder barrel. Polishing took place before installation in the chamber.
\begin{table}[htdp]
\caption{Set of parameters to study scaling laws}
\begin{center}
\begin{tabular}{|l|l|l|l|l|}
\hline
No. & Diameter & Distance & Pressure range & Scaling factor \\
    & mm       & mm       & hPa		   & \\
\hline
III) & $4.3$ & $4.0$ & $200-1000$ & $a\approx2.0$	\\
IV) & $3.0$ & $2.8$ & $290-1430$ & $a\approx1.4$	\\
V) & $2.1$ & $2.0$ & $400-1600$ & $a=1$		\\
\hline
\end{tabular}
\end{center}
\label{tab:parameters_scaling}
\end{table}

\begin{figure}[ht]
	\centering
	\includegraphics[scale=1.00]{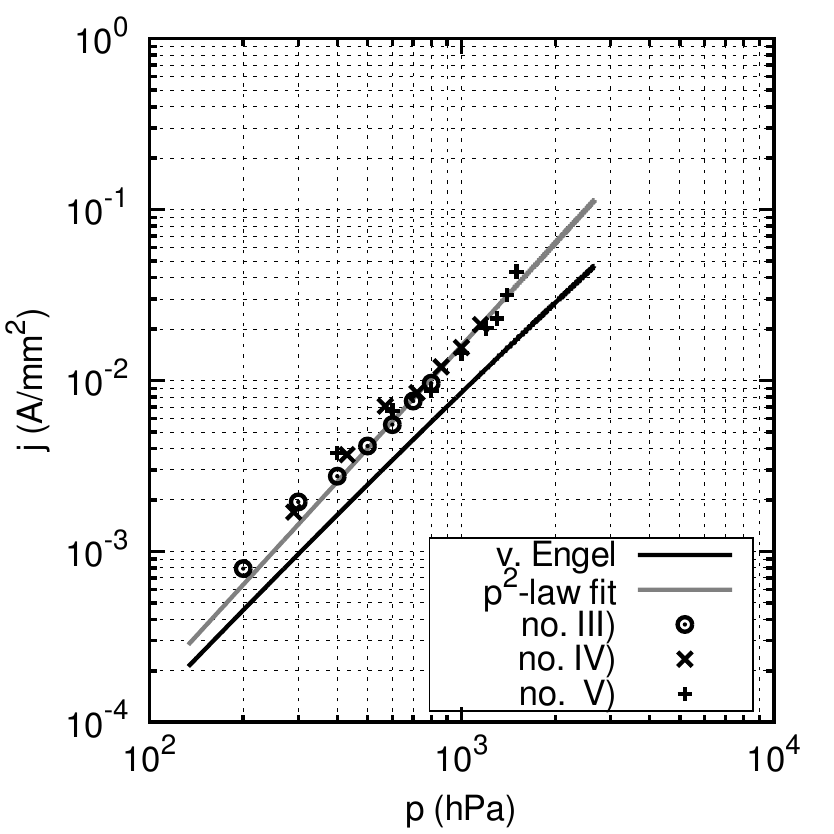}
	\caption{Normal current densities for three similar discharges (symbols) according to Tab.~\ref{tab:parameters_scaling}. The solid gray line is the p$^2$-law with a reduced normal current density of $1.597\times 10^{-8}\;$A/(hPa$^2$mm$^2$). A. von Engels modified version of the law \cite{Engel1933} is depicted as black solid line.}
	\label{fig:joverp2}
\end{figure}

For each measurement the current was increased until the whole cathode surface was covered by the negative glow. At this transition point from a normal to an anormal glow discharge the current value was recorded (Fig.~\ref{fig:joverp2}). Due to constructional  limitations it was not possible to take pictures of the negative glow covering the cathode surface. The fuzzy appereance of the negative glow together with an increased light intensity at elevated pressures tricks the eye easily. This approach is also the biggest source of error for these measurements. 

Every series was fitted with the p$^2$-law. The arithmetic mean of all three results give an estimate for the reduced normal current density
\begin{equation}
\Big(\frac{j}{p^2} \Big)_n=j_n=1.597\times 10^{-8}\pm1.846\times 10^{-10}\,\frac{\textrm{A}}{\textrm{hPa$^2$mm$^2$}}.
\label{equ:p2law}
\end{equation}
This value is to the same order of magnitude as for the values in \cite{Raizer1991}, p.~183 for Fe and Ni (each with $1.238\times 10^{-8}\;$A/(hPa$^2$mm$^2$) at room temperature ($285\;$K).
\subsubsection{Discussion}
The distribution of data points in Fig.~\ref{fig:joverp2} shows a slight s-shaped deviation from the straight gray line given by the quadratic similarity law at elevated pressures around $p\approx 600\;$hPa. For lower pressures the data points are above the fit and for higher pressure they are on the fit.\\

A. von Engel et al. accounted this change to an increase in temperature and hence reduction in density \cite{Engel1933}. In order to add a density correction to the $p^2$-law (Fig.~\ref{fig:joverp2}, black line) he used the ideal gas equation and the average temperature in the cathode fall region $\overline{T}$ (Eq.~\ref{equ:tbar})
\begin{equation}
j=j_n\Bigg (\frac{pT_0}{\overline T} \Bigg )^2.
\end{equation}
The normal temperature $T_0$ and the reduced normal current density $j_n$ are tabulated in Tab.~\ref{tab:tbar_param}. According to their calculations the exponent of the pressure $p$ converges from $2$ to $4/3$. The mathematical results are slightly below the measured values but in reasonable good agreement taking the uncertainty of the measurement method into account. Mezei et al. \cite{Mezei2001} observed a smililar trend for molecular gases. They observed a convergence of the exponent for air to $1/2$. Additionally, a change in the diameter of the positive column was observed. 
\subsection{Gas heating and diameter of the positive column}
\label{sec:dpT}
\begin{figure}[ht]
	\centering
	\includegraphics[scale=1.00]{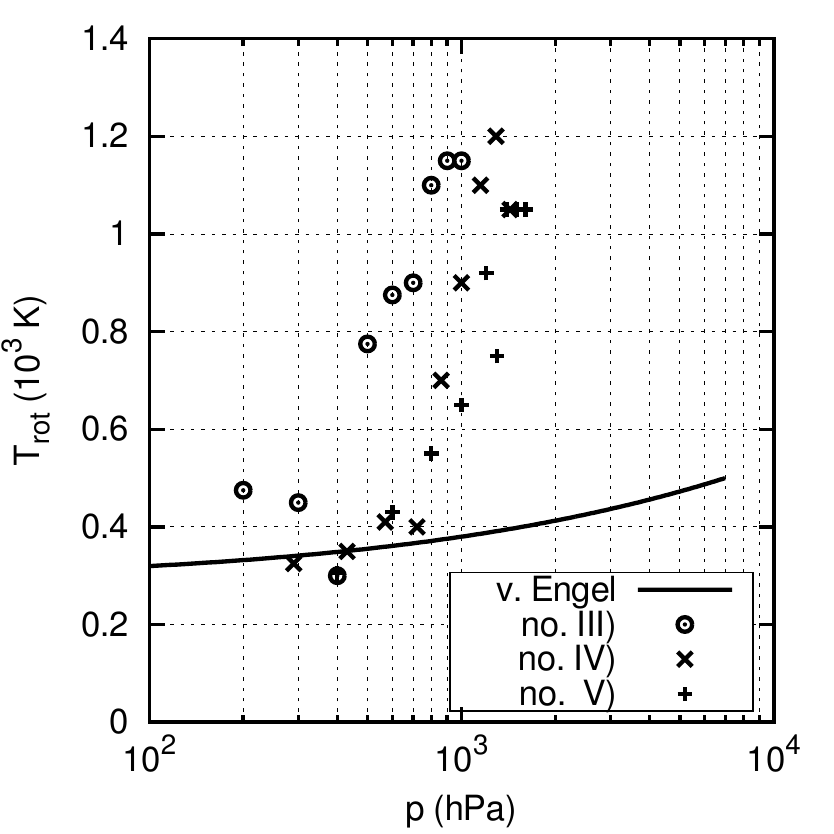}
	\caption{Rotational temperatures from the OH-band for the measurements in Tab.~\ref{tab:parameters_scaling} at the transition from normal to anormal glow. The black solid line is the temperature computed according to Eq.~\ref{equ:tbar}.}
\label{fig:T_over_p}
\end{figure}

To quantify neutral gas heating in the plasma, optical spectra were taken at the transition from normal to anormal glow. OH impurities were always present during these measurements. A fit of the OH-band with Lifbase gave an estimate for the rotational temperatures for all three measurement series (Fig.~\ref{fig:T_over_p}). The rotational temperature is below the electron temperature and gives an upper boundary for the gas kinetic temperatures.

\subsubsection{Discussion}
Temperatures range from room temperatures up to $1200$~K and are an estimate for the temperature in the cathode fall region. Although the emission of the whole plasma is recorded this assumption is justified since the negative glow is far brighter than the positive column. Whereas the measurement series no.~IV) and no.~V) overlap, the data points of no.~III) show higher temperatures at lower pressures. It is believed that the reason for this behavior is due to lower light intensities of series no.~III), a worse signal to noise ratio and hence a bigger uncertainty in the Lifbase fits. Furthermore, the OH-band is only visible as continuum with the used spectrometer which also causes uncertainties.\\
The plasma show strong variations in brightness, overall optical appearance and dimensions with increasing pressure. At some points self-organized structures at the anode, which are reported for this study in \cite{Maszl2011b} and striations can be observed. These effects are known for a long time for high \cite{Ammelt1998} and low pressures and beautifully documented with handmade colored drawings in \cite{Lehmann1901}.\\

An attempt to apply the model from Eq.~\ref{equ:tbar} with the parameters summarized in Tab.~\ref{tab:tbar_param} is only successful for relative low pressures (Fig.~\ref{fig:T_over_p}, black line). The necessary constants were found in \cite{Raizer1991}, p.~182-183 and converted to SI-units. 
\begin{table}[ht]
	\centering
	\begin{tabular}{|c|c|c|c|c|c|}
		\hline
		$V_C$	& $\alpha_G$	  & $d_n$	& $j_n$		& $T_0$ & $T_E$ \\
		V	& W/(mm$\;$K$^2$) & mm$\;$hPa	& A/(mm$^2$hPa$^2$) & K & K\\
		\hline
		$177$	& $5.3\times 10^{-7}$ & $14.3$ 	& $1.597\times 10^{-8}$ & $285$ & $288$ \\
		\hline
	\end{tabular}
	\caption{Cathode fall voltage $V_C$, $\alpha_G$ is the thermal conductivity $\lambda$ of He over $T_0$,  reduced normal cathode fall thickess $d_n=pd_C$, reduced normal current density $j_n=(j/p^2)_n$, temperature at normal conditions $T_0$ and temperature of the water-cooled electrodes $T_E$, which are used in Eq.~\ref{equ:tbar}. All parameters from \cite{Raizer1991}, p.~182-183.}
	\label{tab:tbar_param}
\end{table}

At elevated pressures the measured rotational temperatures rise very strongly and the model does not fit the data at all. One explanation for this behaviour is that the diameter of the positive column shows a pressure dependence \cite{Mezei2001}. At elevated pressures the diameter of the discharge decreases. In order to quantify this effect and to obtain a pressure depending scaling factor $f(p)$, gray value profiles were taken from pictures of the positive column. The full width half maximum is then an estimate for the diameter. For pressures above $290\;$hPa the diameter decreases almost linearly. Therefore the scaling factor is normalized to the diameter $D_{290}$ of the discharge at $290\;$hPa
\begin{equation}
f(p)=\frac{D}{D_{290}}=1+\frac{1}{6}\Bigg (1-\frac{p}{290\,\textrm{hPa}}\Bigg ).
\end{equation}
Striations and other self organization phenomena in the positive column give rise to huge uncertainties in the diameter measurement.\\
The positive column follows a path of higher temperature and hence lower resistivity which is also reported for plasma on chip devices for example \cite{Eijkel2000}. This means also that power is dissipated in a smaller volume which again increases the temperature further. High temperatures were also reported for pin to plate discharges in air \cite{Staack2005} for example. Other works emphasize the important role of gas heating in shaping the electric field in the negative glow and the anode layer \cite{Mezei2001,Wang2006}. Therefore the obtained experimental temperature results in Fig.~\ref{fig:T_over_p} appear to be reasonable. A proper fit should take the reduction of volume and the scaling $f(p)$ of the positive column into account.
\subsection{Reduced normal electric field in the positive column}
\label{sec:Ep}
\begin{figure}[ht]
	\centering
	\includegraphics[scale=1.00]{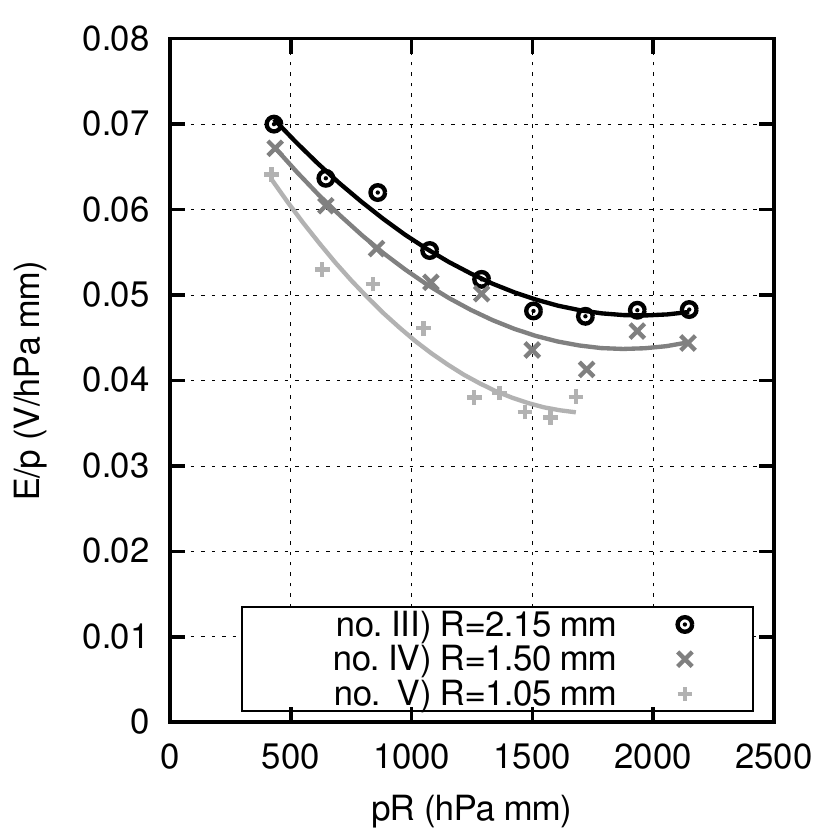}
	\caption{Representation A): Reduced field strength in the positive column at $I_D=25\;$mA for three different electrode diameters (Tab.~\ref{tab:parameters_scaling}). $R$ in this representation is the constant electrode radius. The data points are fitted with a $2^\textrm{nd}$ order polynom.}
	\label{fig:Eop_over_pR}
\end{figure}

The field strength in the positive column is approximately constant. An increase of the electrode  distance $d$ causes therefore a linear increase in the discharge voltage $V_D$, if the discharge current $I_D$ is kept constant  (Fig.~\ref{fig:setup_profile}b). Using the described method, column gradients were measured for the three setups from Tab.~\ref{tab:parameters_scaling} at a constant discharge current of $I_D=25\;$mA. For all measurements the linear slope decreased a little bit between $d=3-4\;$mm. At this point the aspect ratio $d/D$ becomes unfavorable. The field strength was computed therefore below this kink.\\
In principle two ways of representing the obtained data is possible. A) The reduced field strength is plotted over the product of the pressure in the chamber and the radius of the electrode radius (Fig.~\ref{fig:Eop_over_pR}).

\begin{figure}[ht]
	\centering
	\includegraphics[scale=1.00]{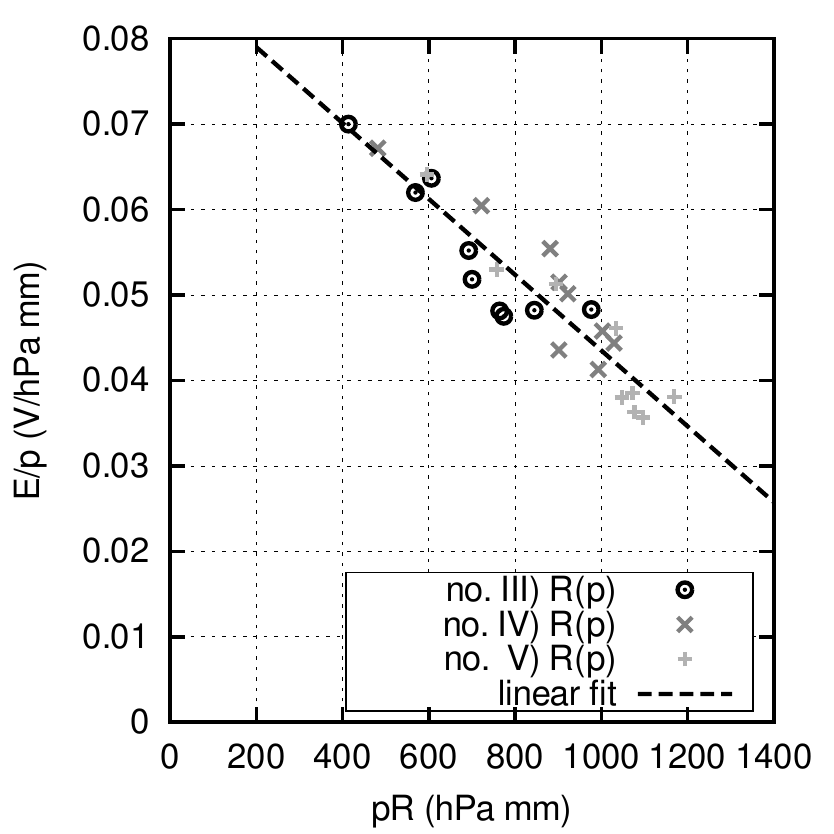}
	\caption{Representation B): Reduced field strength in the positive column at $I_D=25\;$mA for three different electrode diameters (Tab.~\ref{tab:parameters_scaling}). $R$ in this representation is the pressure dependent diameter of the positive column. The solid line is a fit of all data points.}
	\label{fig:Eop_over_pR(p)}
\end{figure}

B) Instead of the constant electrode diameter, the radius of the positive column is taken into account. This radius is not constant anymore since it depends on the discharge conditions. At maximum electrode separation a picture was taken to infer the radius of the positive column. The full width half maximum of the positive columns gray values profiles were taken as measures for the diameters. These radii information were used to compute $pR$ instead of the constants electrode radius (Fig.~\ref{fig:Eop_over_pR(p)}). 

For representation A) (Fig.~\ref{fig:Eop_over_pR}) the obtained data points show no overlap but a monotonic decrease. A quadratic polynom fit is possible for each data set. Taking the decrease of the positive column diameter into account B) (Fig.~\ref{fig:Eop_over_pR(p)}) the data points are scattered but the overall trend becomes linear and the data points overlap.
\subsubsection{Discussion}
One problem here is the proper representation of the data. Representation A) uses the constant electrode radius (Fig.~\ref{fig:Eop_over_pR}) whereas representation B) takes the diameter of the positive column into account (Fig.~\ref{fig:Eop_over_pR(p)}).\\
The field strength in the positive column is among other things determined by the loss mechanisms. At low pressures the plasma is usually in contact with the chamber wall (e.g. glass tube) and radial ambipolar diffusion is the main loss mechanism. For higher pressures recombination becomes important \cite{Chen1984} and the plasma is often constricted. For a free-standing electrode to electrode configuration (Fig.~\ref{fig:DC_expsetup}) like in the presented study, the plasma is surrounded by the neutral working gas. Therefore the radius of the plasma is determined by the radial loss mechanisms for charged particles and not the electrode diameters. A measure for the radius $R$ of the positive column in Fig.~\ref{fig:Eop_over_pR} should therefore be the diffusion length $\Lambda$. Since this parameter is not known easily, the diameter of the positve column is a good measure for the loss mechanisms. For the low pressure case, the diameter of the positive column corresponds to the electrode diameter since the plasma is in contact with the wall and the diffusion length in a cylindrical geometry is directly proportial to the radius ($\Lambda=R/2.4$).
We therefore emphasize the importance to take the diameter of the positive column and recommend representation B) (Fig.~\ref{fig:Eop_over_pR(p)}) as proper representation for DC discharges at elevated pressures! \\

The used setup free-standing electrode to electrode configuration (Fig.~\ref{fig:DC_expsetup}) and the produced plasma surrounded by neutral gas. Results depicted in \cite{Raizer1991}, p.~196 imply a constant value of $E/p=0.08\;$V/(hPa$\;mm$) for $pR>105\;$hPa$\;$mm and a discharge current of $I_D=25\;$mA in the low pressure case. The obtained results in Fig.~\ref{fig:Eop_over_pR(p)} are to the same order of magnitude but show an approximately linear decrease. In contrast to the representation with a constant $R$ the data points overlap indicating that the scaling holds in principle although $E/p$ is not constant. Following a similar argumentation like for the reduced normal current density the decrease of $E/p$ can be caused by thermal heating. The rotational temperatures are in the range between $300\to 1000\;$K for discharge powers between $6\to10\;$W.

\section{Implications on pulsed high power plasmas}
In the previous section we demonstrated how similarity/scaling laws can be used to scale experiments. We demonstrated with a lab scale setup that these laws are far from being absolute, even for a simple system like a DC glow discharge. As described in the introductory part Sec.~\ref{sec:scalinglaws} physical processes can violate these laws and also thermal heating can have a significant impact on the discharge performance and behavior. Reliable predictions are therefore only possible in certain ranges of applicability. In this section we will highlight the added complexity of non-stationary high power plasmas like High Power Impulse Magnetron Sputtering (HiPIMS) in contrast to DC plasmas. The discussion will be limited to magnetron sputtering of metals with argon. In order to keep the discussion simple reactive magnetron sputtering will not be covered.\\

\begin{figure}[ht]
	\centering
	\includegraphics[scale=1.00]{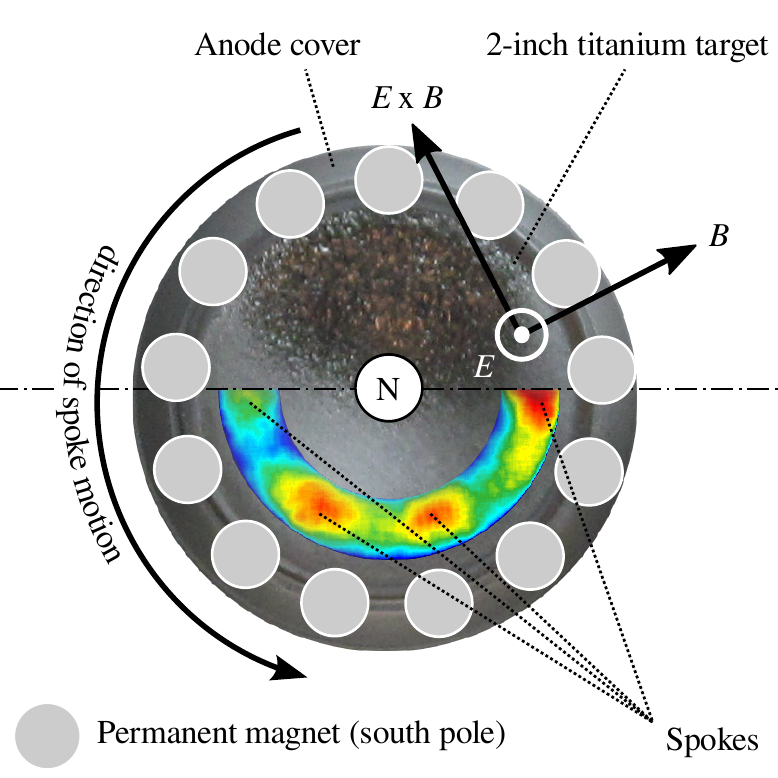}
	\caption{2-inch magnetron cathode with a titanium target. The magnetic field is created by thirteen outer magnets with south pole up (gray circles) and a center magnet with north pole up (white circle). A part of the racetrack plasma in the spoke regime is presented in false colors.}
	\label{fig:magnetron}
\end{figure}

In contrast to the presented glow discharge, in HiPIMS the cathode is a magnetron with typical magnetic fields in the range of 100~mT (Fig.~\ref{fig:magnetron}) and operated as sputter source for deposition. The pressure is lower and to the order of $5\times 10^{-3}$~hPa which gives a mean free path between collisions in the range of several cm.

The cathode is operated in a pulsed mode. During a typical pulse of few tens of $\mu$s, peak power densities of several kW/cm$^{-2}$ are reached on the target producing very high plasma densities and a high ionization degree. At highest powers the plasma  (Fig.~\ref{fig:magnetron}) often shows self-organization, symmetry breaks and hence the formation of so-called spokes. These are described in Sec.~\ref{sec:spokes} in more detail. Furthermore, during the pulses an energetic ion flux arrives at the substrate. The contribution of energetic ions to the growth flux improves thin film properties and distinguishes HiPIMS plasmas from conventional Direct Current Magnetron Sputtering (dcMS). Duty cycles of only a few percent or less are permissible to limit the thermal load on the target to ensure its structural integrity. HiPIMS is nowadays a promising technique for numerous applications \cite{Sarakinos2010}.\\

At first we want to compare the parameters of the reduced current density and reduced field strength of a DC discharges with a HiPIMS plasma. Typical peak discharge currents of a 2-inch magnetron at 0.5~Pa and a titanium target are to the order of 100~A. If one assumes that every spoke constitutes a single DC discharge Hecimovic et al. \cite{Hecimovic2015} find an approximate current density within the spoke of 40~A/cm$^2$ and hence a reduced current density of $(j/p^2)_n=1.6\times 10^3\,$A/(hPa$^2$mm$^2$). This value is several orders of magnitude higher if compared to DC discharges (Equ.~\ref{equ:p2law}). There are indications that spokes are surrounded by a potential hump with a height of approximately 25~V \cite{Maszl2014,Anders2013}. If one assumes a gradient length of 10~mm and a pressure of 0.5~Pa the reduced field strength computes to $E/p=500\,$V/(hPa$\,$mm). Also this value show a significant difference to classical DC plasmas (Fig.~\ref{fig:Eop_over_pR(p)}) with values around 0.05~V/(hPa\,mm).
\subsection{Ignition of pulsed discharges}
Already in the early days of plasma physics differences in the ignition behavior of stationary and voltage impulses were documented \cite{Rogowski1926}. Also the influence of space charges were discussed \cite{Rogowski1931}. In HiPIMS, ignition depends on the magnetic field, the slew rate of the voltage and also on the pulse off time. The onset delay of the discharge current can be separated in a statistical and formative time lag \cite{Kudrle1999,Anders2011}. If a significant background ionization from an previous pulse is left, breakdown delay times are significantly reduced.\\
\subsection{Plasma chemistry and gas heating}
During the pulses the electron density increases and reaches values up to $10^{19}\;$m$^{-3}$ \cite{Alami2005}. But not only the absolute values but also the density distribution is changing with time \cite{Bohlmark2005}. Other quantities like the plasma potential shows also strong variations throughout the pulses \cite{Mishra2010,Mishra2011,Rauch2012,Liebig2013}.\\
Depending on the pulse length the plasma chemistry can change significantly. For short pulses ($\sim 5\;\mu$s) the discharge is sputter gas dominated whereas for longer pulses ($\sim 20\;\mu$s) self-sputtering becomes dominant \cite{Konstantinidis2006}. This change can also be observed in spatial emissivity maps of the different species \cite{Hecimovic2012}. The reason for the strong rarefaction of the sputter gas is a sputter wind which originates from sputtered metallic species \cite{Hoffman1985,Rossnagel1998,Palmucci2013}.\\
Similarly to DC plasmas, gas heating plays an important role on the density of the working gas. Time-resolved tunable diode-laser absorption spectroscopy on argon metastables measures temperatures up to 1600~K \cite{Vitelaru2012}. Also the cathode cannot be considered as cold. Despite water cooling the plasma heats a titanium target up to a surface temperatue of 870$^\circ$C measured by IR radiation \cite{Cormier2013}. It is also likely that during short high power pulses glowing hot spots on the target are created. In these areas, thermionic electron emission will occur with a significant higher electron production rate than electrons released by kinetic or potential emission in cold regions.
\subsection{Electron and ion energy distribution functions}
Electrons in HiPIMS cannot be assumed to be single maxwellian. Three distinct groups of electrons are observed which are described as ''super-thermal'', ''hot'' and ''cold'' with effective temperatures of $70-100\,$eV, $5-7\,$eV and $0.8-1\,$eV \cite{Poolcharuansin2010}. The ''super-thermal'' population are secondary electrons which are accelerated in the cathode fall. According to an analytical model these electrons are cooled by inelastic ionization and excitation collisions with neutral species, and by elastic Coulomb interactions with the cold Maxwellian electron population \cite{Gallian2015}.\\
One important aspect of HiPIMS is the natural occurence of high energetic ions of sputtered species \cite{Bohlmark2006}. For Ti$^{1+}$ the maximum of the cold population is around one eV and the maximum of the hot population around $20\,$eV \cite{Maszl2014}. The hot ions are believed to be responsible for the excellent mechanical thin film properties which are achieved by the HiPIMS technique. Recent studies show that the formation of internal potential structures during ''spokes'' is likely to be responsible for the acceleration of the ions \cite{Anders2013,Maszl2014}.
\subsection{Heavy particle transport in HiPIMS}
One problem of HiPIMS is the reduced deposition rate compared to direct current magnetron sputtering (dcMS) at the same averaged powers \cite{Gudmundsson2012,Mitschker2013}. This hampers the succesful commercialization of this promising technique. A loss of deposition rate is attributed to the onset of self-sputtering \cite{Alami2006}. The applied discharge power also has a significant impcat on the ionized metal part fraction of the growth flux which is reduced with increasing power \cite{Poolcharuansin2012a}. The proposed mechanism is a combination of the ionization probability of the sputtered species and the return effect of the ions. But also the magnetic field is an important parameter which controls the downstream plasma transport. Experiments show that for high magnetic fields a larger pre-sheath potential drop is created which causes an inefficient ion extraction \cite{Meng2014}.
\subsection{Magnetic fields and internal currents}
HiPIMS cathodes are operated in a $E\times B$ configuration which give rise to the internal Hall current and the formation of a dense plasma above the racetrack. Three different types of internal current systems are identified \cite{Lundin2011}. An interesting aspect is that cross-field electron transport cannot be described by classical Bohm diffusion. These authors present that the axial current is too high in comparison with azimuthal current in a classical picture and speculate that an anomalous transport mechanism has to be present \cite{Brenning2009} like a modified two stream instability \cite{Lundin2008a}. The influence of internal currents on the static magnetic field is demonstrated by  \cite{Bohlmark2004}.\\
Also the current density distribution over the racetrack is subject to variations \cite{Clarke2009,Poolcharuansin2015}. Especially, in the spoke regime one assumes that a significant part of the discharge current is conducted through the ionization zones. The differences in sputter yield below and in between spokes will also give rise to variations in plasma chemistry.
\subsection{Symmetry breaks and travelling ionization zones}
\label{sec:spokes}
Travelling ionization zones or so-called ''spokes'' are observed by several authors \cite{Ehiasarian2012,Winter2013,Brenning2013}. In the spokes regime the homogeneous plasma torus known from dcMS plasmas changes to a finite number of plasmoids which are rotating over the racetrack (Fig.~\ref{fig:magnetron}). This pattern is usually regular with a certain number plasmoids. The number of plasmoids is referred to as quasi-mode number. In our case the mode number is typically higher than in the literature since our magnetic field has strong ripple in azimuthal direction (Fig.~\ref{fig:magnetron}). Magnetrons with ring magnets usually show lower mode numbers. These highly localized, self-organized structures, with a quasi-mode number typically between one and four for a 2-inch magnetron, rotate in the $E\times B$ direction in front of the target surface. The velocity of these spokes is ten times smaller than the $E\times B$ drift according to single particle motion. Quasi-mode number changes and spoke merging was observed directly by probe array measurements \cite{Hecimovic2015}. At low powers or dcMS-like conditions also motions against the $E\times B$ direction are observed \cite{Ito2015}. The generation mechanism of the spokes and their influence
on the transport properties are not fully understood yet. During spokes also plasma flares can be observed \cite{Ni2012}. Here, hot plasma is ejected towards the substrate above the spoke. These flares indicate the presence of azimuthal electric fields which convect the plasma out via an $E\times B$ process. One ongoing discussion is the role of secondary electrons in the discharge. Some authors connect the shape of the spokes with these electrons \cite{Hecimovic2014} which are also to be thought as main energy source. Other authors question the role of secondaries. Dedicated models show that ohmic heating can dominate over secondary emitted electron energization in the cathode sheath \cite{Huo2013} but still need experimental validation. 
\subsection{Discussion}
Because of the transient nature of HiPIMS plasma it is unlikely to find simple scaling laws like Eq.~\ref{equ:scaling_laws}. But how is it possible to define such a plasma unambigously with regard to knowledge based thin film design? Alami et al. \cite{Alami2006} found four different regimes. In an IV representation of the current and voltage pulse these regimes are distinguished by their differential resistance. These regimes can also be correlated with the optical appereance of the torus in terms of spoke formation and deposition rate. In principle this approach can give a marker or ''fingerprint'' of a certain plasma if it is universal \cite{Arcos2013,Arcos2014}. HiPIMS plasma are like almost all technological plasma bounded plasma systems \cite{Kuhn1994}. The evolution of the discharge is highly influenced by the performance of the vacuum and gas feed system \cite{Arcos2014}. Furthermore, the influx of neutral argon in the discharge zone could also have a significant impact on spoke formation and evolution if one takes ''predator and prey'' models into account \cite{Gallian2013,Hecimovic2015}. Here, the high electron density in the spokes is responsible for the rarefaction of the noble gas. Therefore, the spoke moves towards higher argon densities. The whole dynamics of the system relies on how fast argon can be refilled in the wake of the spoke.\\
But also the power supply has a significant influence on the discharge evolution. HiPIMS power supplies can be typically operated in voltage (V-) or average power controlled P-mode. In the first case the controller tries to keep the pulse voltage constant. Although preferable for experiments a thermal instability can cause a continous growth of the peak current until the discharge starts to arc. A fast controller in combination with a big capacitor bank will keep the voltage constant throughout the pulses and the regimes described above  cannot be observed. In P-mode the discharge is stable but the voltage is subject to changes from pulse to pulse which impedes the comparability. Aside from the controller, internal or parasitic inductances $L$ have also a strong influence since they limit the current rise rate for a given voltage $\textrm{d} i/\textrm{d} t=U/L$. Teresa de los Arcos et al. installed a passive low-pass filter to limit the current rise time. This avoids the often observed initial current peak \cite{Anders2011}, therefore hot spots and thermionic electron emission on the target and allows to reach overall higher power densities. Recent power supplies even start with a higher voltage pulse in the beginning to drive the discharge as fast as possible to the desired maximum current. The voltage is then lowered to reach a stable plateau. Summarizing, the transient nature of the discharge, the changes in plasma chemistry, the appereance of symmetry breaks and self-organization, the technological boundaries and active controls in the power supplies impede to find an unambigous set of parameters or an universal IV finger print to specify a certain discharge unambigous.\\

Huge efforts are therefore spend on the modelling side. If reliable plasma models and thin film growth models can be coupled, knowledge based thin film design could be achieved. Because of the huge complexity multiple models for the plasma itself have to be coupled. First attempts to model the neutral gas dynamics without plasma were done by \cite{Bobzin2013}. They attempt to couple it to plasma models in the future. Phenomenological equilibrium models allow the computation of ioniztaion fraction of sputtered species \cite{Vlcek2010}. Global models give highly welcome insight in fluxes of species during the pulses which cannot easily obtained by experiments \cite{Raadu2011,Kozak2011,Kozak2012,Huo2014}. Especially plasma parameters above the racetrack are valuable because diagnostic possibilities are limited in that region. Particle In Cell (PIC) codes can help to shed more light on this quantities \cite{Minea2014}. Taking the important role of spokes into account future work will also have to adress them. First attempts were made with a phenomenolgical model \cite{Gallian2013} but in order to get more insight in the underlying physics eventually three dimensional time depending models will be required. 

\section{Conclusion}
A brief introduction to scaling laws for DC glow discharges was given and it was demonstrated, how they can be utilized to dimension normal glow discharges. Further a  high pressure discharge chamber was used for experimental studies. Measurements of breakdown voltages give the result that the pressure $p$ and the inter-electrode $d$ cannot be varied independently. Both curves start to deviate at $pR\approx 3000\;$hPa$\;$mm. The semi-empirical Paschen law overestimates the breakdown voltage in that range. A possible reason can be field amplification at the electrode edges and therefore a reduced breakdown voltage. Furthermore, the fit and hence the fit parameters of the first Townsend coefficient $\alpha$ are only valid for certain ranges.\\
The normal current density shows a deviation from the $p^2$-law for pressures above $p\approx 600$~hPa. Other authors account this change to an increase in temperature and the reduction in neutral gas density. Temperature estimates were measured via OH impurities and give maximum rotational temperatures up to $1200\;$K for pressures around $1000\;$hPa. Finally the field strength in the positive column was measured via column gradients for a constant current $I_D=25\;$mA. Again the linear decrease was motivated by an increase in the temperature. Further, it was discussed why the diffusion length or the width of the positive column in contrary to the electrode diameter should be used to represent the data in the high pressure regime.\\
In the last part of this article we briefly reviewed the physics of HiPIMS plasmas and presented their complexity. The comparison with DC plasmas shows that look-up tables or simple IV-characteristics will not be sufficient to define a certain plasma unambigously. For these kind of systems complex models are a necessity which have to be able to translate the manner of how the power is supplied, to plasma parameters and further to thin film properties. Only with sophisticated models the goal of knowledge based thin film design can be reached in the future. Fortunately, big efforts are spend by modellers in the community to gain more insight in the underlying physics of regions, which are not easily accessible by diagnostics, of these highly interesting plasmas.

\section*{Acknowledgements}
This project is supported by the DFG (German
Science Foundation) within the framework of the Coordinated
Research Center SFB-TR 87 and the Research Department ''Plasmas
with Complex Interactions'' at Ruhr-University Bochum. 

\section*{References}
\bibliography{bibmac} 

\end{document}